\pdfoutput=1
\documentclass[aps,pre,twocolumn,amsmath,amssymb,notitlepage,superscriptaddress,longbibliography,nofootinbib,10pt]{revtex4-1}
\usepackage[T1]{fontenc}
\usepackage[left]{lineno}
\usepackage{amsmath}
\usepackage{amssymb}
\usepackage{bm}
\usepackage{graphicx}
\usepackage{xcolor}
\definecolor{bleuf}{rgb}{0,0,1}
\definecolor{redf}{rgb}{0.61,0.24,0.38}
\usepackage[%
  colorlinks=true,
  urlcolor=redf,
  linkcolor=redf,
  citecolor=redf
]{hyperref}
\usepackage{url}
\usepackage{soul}
\usepackage{upgreek}
\usepackage[normalem]{ulem}
\usepackage{float}
\usepackage{appendix}

\begin{document}

\title{Collective self-sorting on a chip}

\author{Emanuel F.~Teixeira}
\email{teixeira@physics.leidenuniv.nl}
\affiliation{Huygens-Kamerlingh Onnes Laboratory, Universiteit Leiden,
PO Box 9504, 2300 RA Leiden, The Netherlands}

\author{Thieu van den Bergh}
\affiliation{Huygens-Kamerlingh Onnes Laboratory, Universiteit Leiden,
PO Box 9504, 2300 RA Leiden, The Netherlands}

\author{Arjen Klok}
\affiliation{Huygens-Kamerlingh Onnes Laboratory, Universiteit Leiden,
PO Box 9504, 2300 RA Leiden, The Netherlands}

\author{Tijn Heesakkers}
\affiliation{Huygens-Kamerlingh Onnes Laboratory, Universiteit Leiden,
PO Box 9504, 2300 RA Leiden, The Netherlands}

\author{Alexandre Morin}
\email{morin@physics.leidenuniv.nl}
\affiliation{Huygens-Kamerlingh Onnes Laboratory, Universiteit Leiden,
PO Box 9504, 2300 RA Leiden, The Netherlands}

\date{\today}

\begin{abstract}
We harness two established ingredients for collective demixing: differential speed and curvature to create a self-sorting device. In binary mixtures, motility differences drive spontaneous spatial segregation, while confinement geometry determines how rapidly and strongly this demixing develops. Using particle based simulations, we systematically identify the geometrical conditions that promote efficient segregation and use these results to guide the design of a finite sorting architecture. We then translate these physical mechanisms into a sequence of curved microfluidic units that progressively amplify the separation of the two species and direct them toward distinct collection regions. Experiments with binary Quincke-roller mixtures confirm that an initially mixed suspension progressively demixes as it propagates through the device, leading to strong enrichment downstream. Our results demonstrate how collective active demixing can be converted into a functional continuous sorting strategy, providing a route toward autonomous microfluidic separation based on particle motility and confinement geometry.
\end{abstract}

\maketitle

\section{Introduction}

\begin{figure*}[t]
    \centering
    \includegraphics[width=\textwidth]{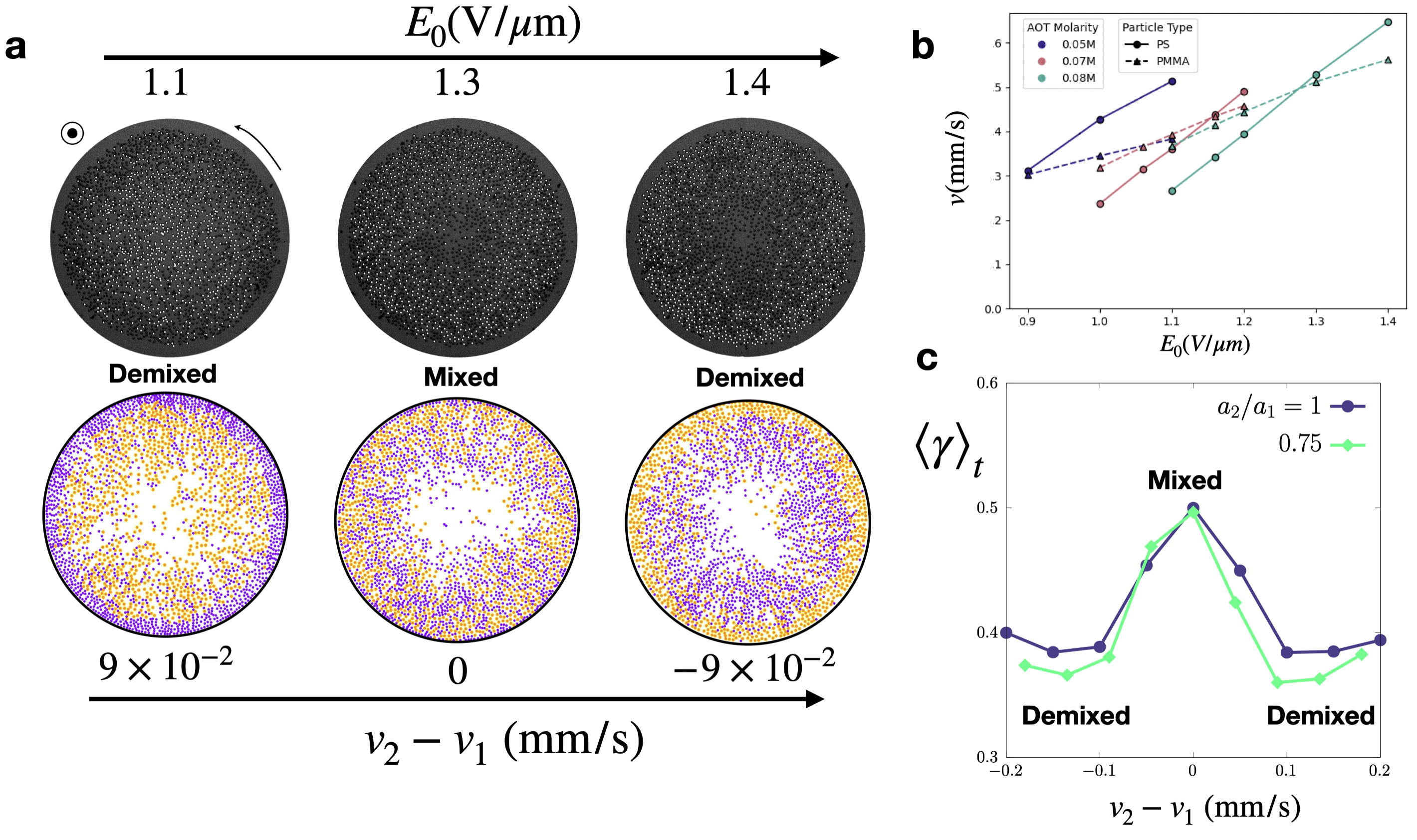}
    \caption{\textbf{Differential self-propulsion speed controls radial demixing.}
$\mathbf{a}$ Experimental observations (top) of binary Quincke-roller flocks confined in a circular well. Reversing the applied electric field switches the relative propulsion speeds of the two species and consequently inverts the demixing pattern: the faster species accumulates near the outer boundary, whereas the slower species is enriched toward the centre.
 Particle-based simulations (bottom) reproduce the experimentally observed radial organization, with faster particles preferentially occupying the outer region and slower particles accumulating in the inner region.
$\mathbf{b}$ Self-propulsion speed as a function of the applied electric field for different AOT molarities for PS and PMMA particles.
$\mathbf{c}$ Demixing parameter as a function of the self-propulsion speed difference for reciprocal and nonreciprocal interactions. Demixing increases with motility contrast in both cases, demonstrating that nonreciprocity is not required for speed-driven demixing.}
    \label{Fig1}
\end{figure*}

Sorting and demixing are central mechanisms by which multicomponent systems acquire spatial organization. In systems evolving at, or close to, equilibrium, thermal fluctuations enable the constituents to explore configuration space, and the resulting organization is ultimately selected by free-energy minimization.  Out-of-equilibrium systems present a distinct challenge. In contrast to systems at equilibrium, their constituents do not necessarily rely on thermal fluctuations to explore configuration space, and the selected state need not minimize a free-energy functional. Segregation can instead emerge from, and be stabilized by, the underlying dynamics. Within this broad class of systems, it is useful to distinguish externally driven mixtures from active ones. In driven systems, spatial reorganization is powered by external actuation or applied fields. A paradigmatic example is the Brazil nut effect, in which mechanical shaking causes granular mixtures to segregate by size~\cite{PhysRevLett.58.1038}. Related principles underlie sedimentation-driven sorting in colloidal suspensions ~\cite{PhysRevLett.74.1478} and phoresis-based mechanisms for particle separation~\cite{lei2022self}. Besides these well know examples, segregation in driven system is still a playground for innovative solutions, such as the stochastic resetting~\cite{cleuren2026sorting}, promising to sort particles by shape.

In contrast with passive systems, active systems provide at the scale of their intrinsic constituents, the mobility to re-organize in space towards a segregated state. A seminal example is the segregation of motile cells with differential speeds ~\cite{JONES1989287}. This was later confirmed by minimal numerical models showing that self-propelled particles with differential speed spontaneously segregate~\cite{PhysRevE.84.031927,mccandlish2012spontaneous}. More broadly, segregation can arise from several forms of phenotypic or dynamical asymmetry. In cellular systems, theoretical and experimental studies have shown that differential adhesion~\cite{steinberg1963reconstruction,graner1992simulation,glazier1993simulation,foty2005differential,PhysRevLett.100.248702,mehes2012collective,mones2015anomalous}, differential contractility ~\cite{harris1976cell,krieg2008tensile,o2016unidirectional,kindberg2021eph,skamrahl2023cellular,PhysRevLett.134.138401}, and differential persistence~\cite{bothe2026contact} can each promote spatial sorting. In passive Brownian mixtures, segregation has likewise been observed to arise from differences in diffusivity~\cite{PhysRevLett.116.058301,PhysRevLett.132.098301}. In synthetic active matter, Maity \textit{et al.} first showed that demixing in mixtures of self-propelled colloidal rollers can emerge from differences in self-propulsion speed, through the combined effects of differential motility, collective motion, and curved confinement~\cite{PhysRevLett.131.178304}. More recently, speed-driven segregation has also been reported in other synthetic active systems~\cite{alvarez2025segregation}. Here, we confirm these predictions experimentally, numerically, and integrate all our findings to design and test a sorting chip based on active collective demixing.
Our new measurements show that a difference in self-propulsion speed is sufficient to drive demixing in binary active mixtures. We demonstrate this using particles made of different materials, considering both equal-sized and differently sized particles. We then compare the experimental measurements with numerical simulations, which quantitatively reproduce the demixing behaviour observed above the flocking transition. We next systematically investigate, both experimentally and numerically, the role of confining boundaries. Consistent with the theoretical description of vortical flocks, our results show that curved boundaries are a key ingredient in promoting and controlling the demixing process. Finally, we combine these insights to design an operational device optimized for autonomous sorting. We conclude by providing an experimental proof of concept for collective active sorting on a chip.

\section{Differential speed drives demixing}
\label{sec:demixing}
Quincke rollers spontaneously organize into polar flocks. Building on the observation by Maity \textit{et al.} that binary Quincke-roller flocks can spontaneously demix when the two species have different motilities, we extend this experimental framework to mixtures with different size ratios and material compositions.
We first examine a binary suspension of fluorescent polystyrene (PS) particles with diameter 
$d_{\rm PS}=10~\mu$m and poly-methyl methacrylate (PMMA) particles with diameter 
$d_{\rm PMMA}=7.5\mu$m, confined in circular wells of radius 
$R=500~\mu$m. The circular well contains $N=2530$ particles, with 
$N_{\rm PS}=1006$ and $N_{\rm PMMA}=1524$. In the microscopy images, the fluorescent PS particles appear as white dots, whereas the PMMA particles appear dark (see Fig.~\ref{Fig1}a). The particles are dispersed in hexadecane containing AOT and loaded into an ITO-coated microfluidic chamber, where a DC electric field applied perpendicular to the glass plates activates Quincke rotation and drives rolling motion along the bottom surface.

In the flocking regime, this dense binary suspension forms a circulating polar flock and undergoes clear radial demixing inside the circular confinement (see Fig.~\ref{Fig1}a). Crucially, the demixing pattern is not fixed by particle size or material alone. Instead, it depends on the applied electric field, which tunes the relative self-propulsion speeds of the two species (see Fig.~\ref{Fig1}b). This makes the PS-PMMA mixture an ideal experimental system to test whether motility difference, rather than size, controls the polarity of active demixing.  

At low field, PMMA particles self-propel faster (Fig.~\ref{Fig1}b) and accumulate near the outer edge of the flock, while PS particles are enriched toward the centre. At higher field, the speed hierarchy is reversed: PS particles become faster and migrate to the outer region, whereas PMMA particles accumulate closer to the centre. Thus, the demixing pattern is inverted by changing the electric field, with the faster species consistently occupying the outer region and the slower species the inner region, as shown in Fig.~\ref{Fig1}a. This field-induced inversion provides direct control over active demixing. By tuning the electric field, we can reverse the demixing polarity and partially modulate the demixing strength. This programmability is essential for device implementation, as it shows that motility-induced demixing can be used as an externally controlled sorting mechanism. 



\section{Particle based simulations}
\label{sec:simulations}
Maity \textit{et al.}~\cite{PhysRevLett.131.178304} introduced a microscopic model for binary mixtures of confined Quincke rollers, incorporating hydrodynamic alignment, electrostatic repulsion, and species-dependent self-propulsion. By coarse-graining these particle-level dynamics, they derived a hydrodynamic theory for the density and polarization fields of the two species. The resulting steady-state density profiles captured the experimentally observed flocking and radial demixing without fitting parameters, identifying propulsion-speed heterogeneity as the primary demixing mechanism and nonreciprocal interactions as a quantitative correction to the spatial profiles.
Here, we directly integrate the same microscopic equations of motion in molecular dynamics simulations of binary mixtures confined within circular wells. This particle-based approach enables a direct comparison with our experimental observations while allowing the speed to be varied independently and extended beyond the experimentally accessible range.

 We describe a binary colloidal flock wherein each roller self-propels at a species-dependent speed, undergoes rotational diffusion, and reorients through hydrodynamic alignment and electrostatic repulsion with neighbouring particles. The position $\mathbf{r}_i$ and orientation $\theta_i$ of particle $i$ evolve as
\begin{align}
    \dot{\mathbf{r}}_i
    &= v_i\,\hat{\mathbf{n}}_i +  \zeta \sum_{j \neq i} \mathbf{F}_{ij},
    \label{eq:translational_dynamics}
    \\
   \dot{\theta}_i
    &= \frac{1}{\tau}
    \frac{\partial}{\partial\theta_i}
    \sum_{i\neq j}H_{\mathrm{eff}}\left(r_{ij},\hat{\mathbf{n}}_i,\hat{\mathbf{n}}_j\right)
    +\sqrt{2D_R}\,\xi_i(t),
    \label{eq:orientational_dynamics}
\end{align}
where $\hat{\mathbf{n}}_i=(\cos\theta_i,\sin\theta_i)$ is the direction of self-propulsion $v_i$ of particle $i$, $D_R$ is rotational diffusivity and $\zeta$ the mobility.
The combined effects of hydrodynamic and electrostatic interactions is described by effective potential 
\begin{equation}
H_{\mathrm{eff}}\left(r_{ij},\hat{\mathbf{n}}_i,\hat{\mathbf{n}}_j\right)
=
A_{ij}(r_{ij})\,
\hat{\mathbf{n}}_i\cdot\hat{\mathbf{n}}_j
+
B_{ij}(r_{ij})\,
\hat{\mathbf{n}}_i\cdot\hat{\mathbf{r}}_{ij},
\end{equation}
where $r_{ij}=|\mathbf{r}_i-\mathbf{r}_j|$ and $\hat{\mathbf{r}}_{ij}=\frac{\mathbf{r}_i-\mathbf{r}_j}{|\mathbf{r}_i-\mathbf{r}_j|}$. The functions $A_{ij}(r_{ij})$ and $B_{ij}(r_{ij})$ quantify, respectively, the alignment and repulsive interactions exerted by particle $j$ on particle $i$. The timescale $\tau$ sets the characteristic relaxation time of the orientational dynamics, while $\xi_i(t)$ is a Gaussian white noise with zero mean and unit variance.
The pairwise force $\mathbf{F}_{ij}$ is an excluded-volume interaction that prevents particle overlap. For interparticle distance $r_{ij}\leq (\sigma_i + \sigma_j)/2$, the repulsive force is $\mathbf{F}_{ij}=-k_c(r_{ij}-(\sigma_i + \sigma_j)/2)\hat{\mathbf{r}}_{ij}$, where $k_c$ is the repulsive stiffness, and $\sigma_i$ and $\sigma_j$ are the diameters of particles $i$ and $j$, respectively.
For a binary mixture of species $\mu$ and $\nu$, let $i$ and $j$ label the interacting particles, and let $\alpha_i,\alpha_j\in{\mu,\nu}$ denote their respective species. When particles $i$ and $j$ interact, the coupling is determined by the matrix element in row $\alpha_i$ and column $\alpha_j$. The interaction matrices are therefore
\begin{align}
    \mathbf{A}(r_{ij})
    &=
    \frac{\mathcal{A}}{r_{ij}^3}\,
    \Theta(r_{ij})
    \begin{pmatrix}
        a_\mu^3 & a_\nu^3 \\
        a_\mu^3 & a_\nu^3
    \end{pmatrix},
    \label{eq:alignment_matrix}
    \\
    \mathbf{B}(r_{ij})
    &=
    \frac{\mathcal{B}}{r_{ij}^4}\,
    \Theta(r_{ij})
    \begin{pmatrix}
        a_\mu^4          & a_\mu a_\nu^3 \\
        a_\nu a_\mu^3    & a_\nu^4
    \end{pmatrix},
    \label{eq:repulsion_matrix}
\end{align}
The four matrix elements correspond to the ordered species pairs $(\mu,\mu)$, $(\mu,\nu)$, $(\nu,\mu)$, and $(\nu,\nu)$. Here, $a_\mu$ and $a_\nu$ are the radii of particles of species $\mu$ and $\nu$, respectively, $\mathcal{A}$ and $\mathcal{B}$ set the alignment and repulsion strengths, and $\Theta(r_{ij})$ restricts the interactions to a finite range. The nonreciprocity arises from the repulsive torque term $\hat{\mathbf{n}}_i\cdot\hat{\mathbf{r}}_{ij}$, combined with the difference in particle size between the two species, which is explicitly encoded in the interaction matrices. Because these matrices are generally not symmetric under the exchange of the interacting species, the microscopic interactions violate action–reaction symmetry.
Details of the numerical methods and parameters, chosen on the basis of experimentally measured values, are provided in the Supplementary Information. Yellow and purple particles are denoted as species 1 and 2, respectively. The diameter of the larger, yellow particles defines the characteristic length scale, such that $a_1=1$, while the diameter of species 2, $a_2$, is expressed relative to $a_1$. All length scales are expressed in units of the radius of species $a_1$.
The propulsion speed of the faster species defines the characteristic velocity scale and is set to $v_{\mathrm{fast}}=1$. Either species may be assigned the larger propulsion speed, allowing us to systematically vary the relative motility of the two species. The rotational diffusion coefficient is fixed at $D_R=0.02$. This value was obtained by converting previously reported experimental measurements for Quincke rollers~\cite{PhysRevLett.131.178304} into the simulation units defined by the characteristic particle diameter and propulsion speed.
 
Simulations of a binary mixture with $a_{2}/a_{1}=0.75$, $N_{1}=1006$ particles of species 1 and $N_{2}=1524$ particles of species 2, confined within a circular well of radius $R=50$, show that a differences in self-propulsion speed is sufficient to induce demixing despite the difference in particle size. Consistent with the experiments, the slower species preferentially occupies the central region, whereas the faster species accumulates near the outer boundary (Fig.~\ref{Fig1}a).
To quantify the degree of demixing, we use the demixing parameter introduced by Belmonte et al.~\cite{PhysRevLett.100.248702}:
\begin{equation}
\gamma = \left\langle \frac{n_{\neq}}{n} \right\rangle,
\end{equation}
where $n_{\neq}$ is the number of neighbouring particles belonging to the other species and $n$ is the total number of neighbours. The brackets denote an average over all particles of both species. Thus, large values of $\gamma$ correspond to a well-mixed state, whereas smaller values indicate stronger demixing.
\begin{figure*}[!t]
    \centering
\includegraphics[width=\textwidth]{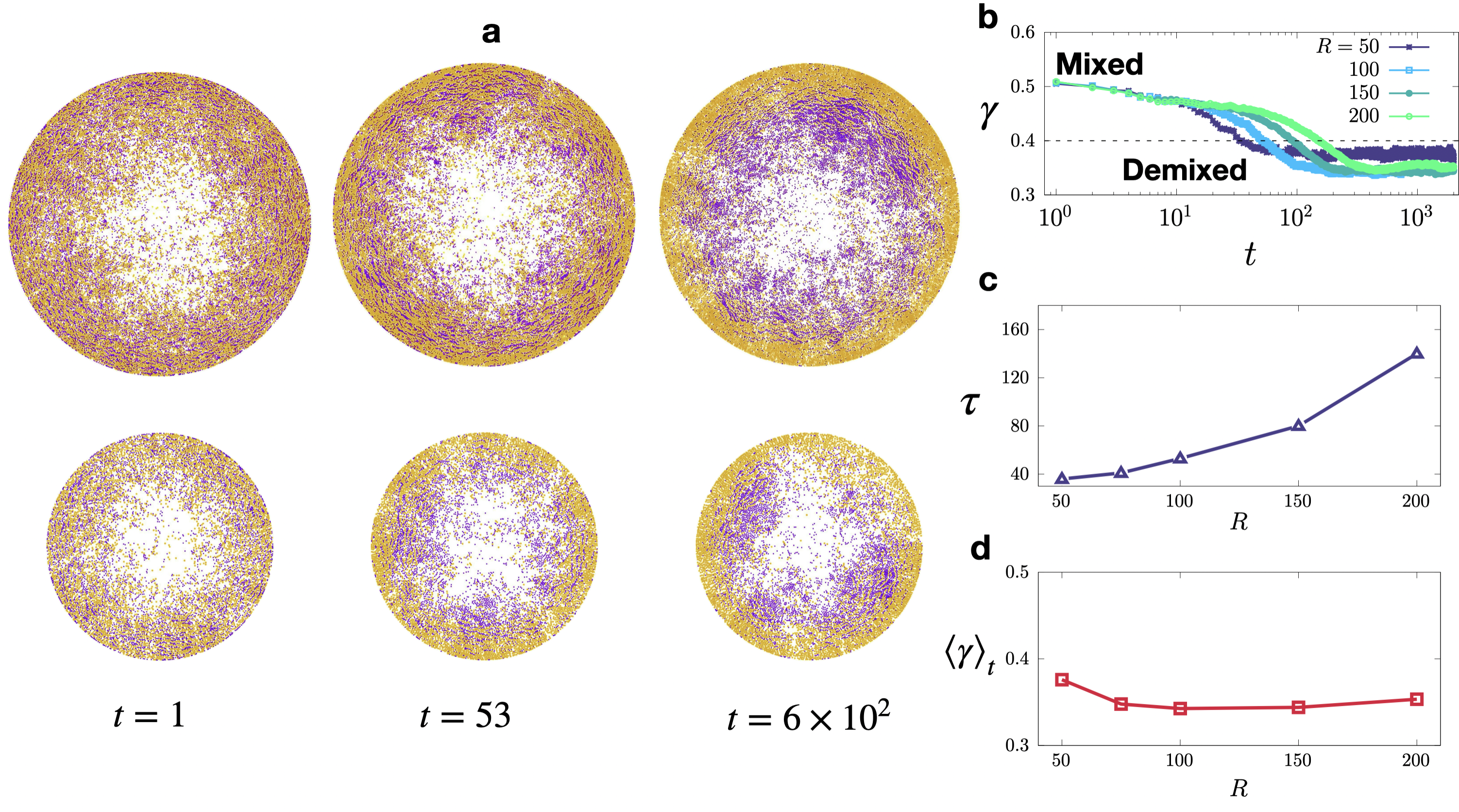}
\caption{\textbf{Smaller confinements accelerate demixing without compromising separation purity.}
$\mathbf{a}$ Configurations of binary active particles confined in circular arenas of different radii, showing the steady-state radial organization of the two species.
$\mathbf{b}$ Time evolution of the demixing parameter, $\gamma$, for different arena sizes. The characteristic demixing time, $\tau$, is defined as the first time at which $\gamma<0.4$, corresponding to the emergence of a macroscopically demixed state with a well-defined radial organization.
$\mathbf{c}$ Characteristic demixing time as a function of the arena radius. Demixing occurs more rapidly in smaller arenas because particles must reorganize over shorter distances to establish the steady-state radial distribution.
$\mathbf{d}$ Steady-state demixing parameter as a function of the arena radius. Its weak dependence on system size demonstrates that reducing the arena size accelerates demixing without substantially decreasing the final separation purity. These results indicate that an array of small arenas can provide a higher processing throughput than a single large arena.}
\label{Fig2}
\end{figure*}
Figure~\ref{Fig1}c shows the steady state value of $\gamma$, averaged over time, as a function of the relative propulsion speed of the two species. Both experiments and numerical simulations, for mixtures of equal and unequal sized particles, show that $\gamma$ is maximal when the two species propel at the same speed and decreases as the speed contrast increases in either direction. Together with the corresponding inversion of the radial density profiles, this non-monotonic dependence demonstrates that the species occupying the outer region is selected by its relative propulsion speed: reversing the speed hierarchy reverses the radial organization. These results establish that a difference in self-propulsion speed is sufficient to drive robust radial demixing under circular confinement.
The simulations further provide a controlled framework for independently probing the effects of confinement geometry and speed difference. A minimal particle based model with species-dependent propulsion speeds therefore captures the essential mechanism underlying the experimentally observed radial demixing and provides a tool for identifying the conditions that maximize sorting performance.
\section{Designing a self-sorting device}
\label{sec:device_design}
\begin{figure*}[t]
\centering
\includegraphics[width=\textwidth]{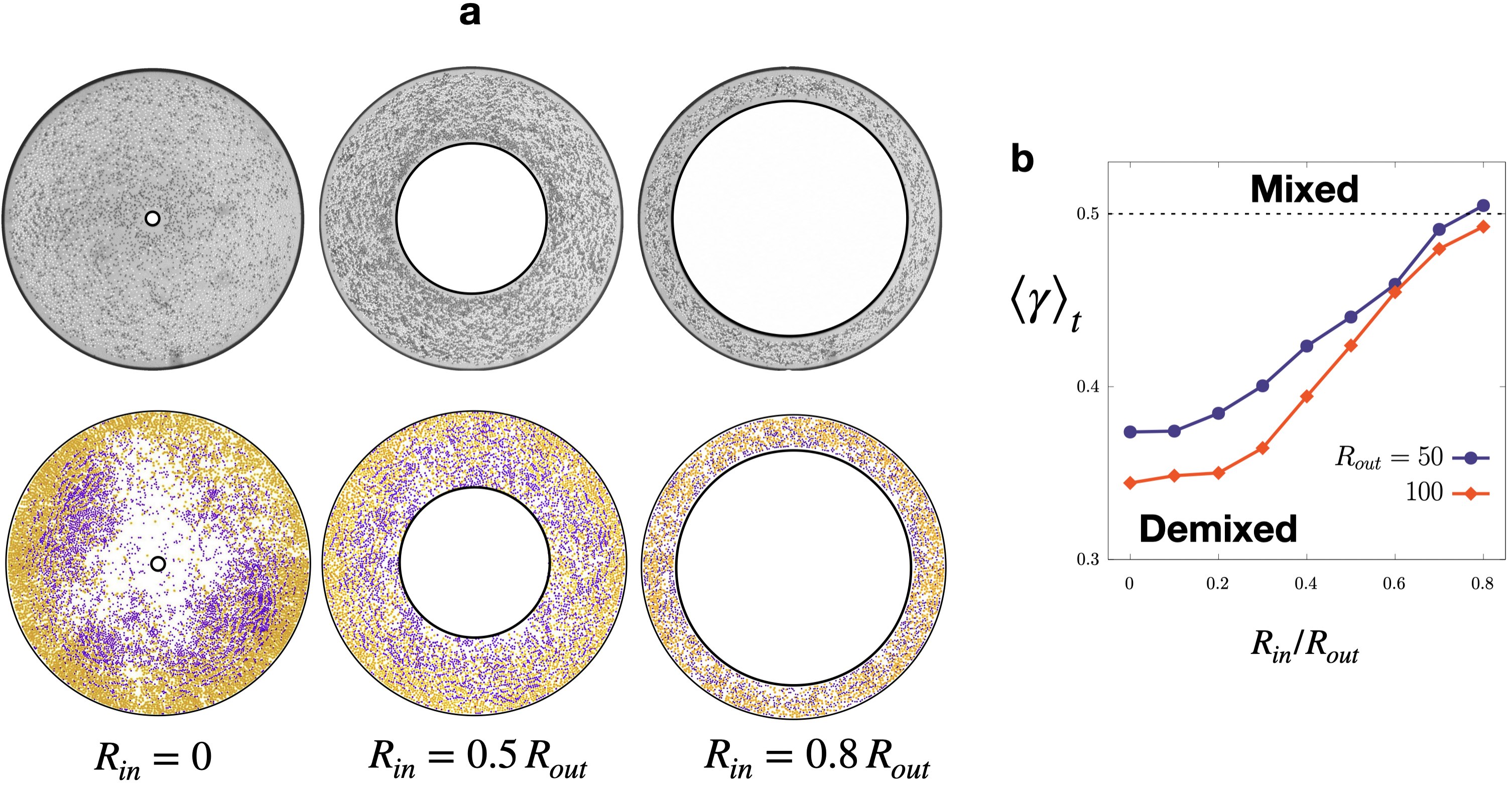}
\caption{\textbf{From arena to circular channels.} \textbf{a.} Experimental snapshots of a binary mixture confined in an annular channel for different values of $R_{\mathrm{in}}/R_{\mathrm{out}}$, showing the progressive loss of radial demixing as the inner radius increases. \textbf{b.} Corresponding simulation snapshots obtained under the same conditions, reproducing the transition from a demixed to a mixed state. \textbf{c.} Steady-state demixing parameter from numerical simulations, $\langle\gamma\rangle_t$, as a function of $R_{\mathrm{in}}/R_{\mathrm{out}}$ for experiments and simulations. Demixing decreases monotonically with increasing $R_{\mathrm{in}}/R_{\mathrm{out}}$, demonstrating that the relative curvature of the inner and outer boundaries controls the efficiency of radial segregation.}
\label{Fig3}
\end{figure*}
Before designing a self-sorting device, we first sought to determine how demixing performance could be optimized. We considered two key criteria: the purity of the separated fractions and the processing throughput. These quantities are governed, respectively, by the demixing level and the time required to reach it.

We therefore investigated the effect of confinement size by performing simulations in circular arenas of different radii. In these simulations, we considered a 50:50 binary mixture with a size ratio $a_2/a_1 = 0.75$, a propulsion-speed ratio $v_2/v_1 = 0.8$, and a packing fraction $\phi = \sum_{i}^{N} \frac{\pi \sigma_{i}^2}{\pi R^2} =0.2$. These parameter values were chosen to be representative of those typically measured in the experiments. In Fig.~\ref{Fig2}a we observe the system in two circular arenas of radius $R = 200$ (top) and $R = 100$ (bottom). The smaller arena reaches the demixed state faster than the larger one, indicating that the demixing timescale depends on the confinement radius. Figure~\ref{Fig2}b shows the evolution of the demixing parameter for different arena radii as the system transitions from a mixed to a demixed state. Although all systems eventually reach a steady demixed state, the time required for the demixing parameter to reach its plateau depends strongly on the arena radius. It is therefore necessary to define a quantitative criterion for determining when the system has reached the demixed state. We define the characteristic demixing time as the time at which $\gamma$ first decreases below $0.4$, a threshold corresponding to the emergence of a macroscopically demixed state with a well-defined radial organization. According to this criterion, we quantify the demixing time and find that demixing proceeds more rapidly in smaller arenas, as evidenced by the decrease in the characteristic demixing time with decreasing arena radius (Fig.~\ref{Fig2}c). This behavior reflects the shorter distances over which particles must reorganize to establish the steady state radial distribution.
Reducing the arena size does not substantially compromise the final degree of demixing. The steady state demixing parameter depends only weakly on the arena size (Fig.~\ref{Fig2}d), indicating that small arenas can retain a separation purity comparable to that of larger systems. Thus, when processing time is a limiting factor, an array of small arenas should provide a higher throughput than a single large arena.
This weak system size dependence differs markedly from that reported by Belmonte~\cite{PhysRevLett.100.248702}, where demixing becomes less pronounced as the system size decreases. In that system, the two components form fully separated macroscopic domains divided by an interface of finite width. The relative contribution of this interfacial region decreases with increasing system size, leading to an apparent enhancement of demixing in larger systems. By contrast, demixing in binary flocks is not driven by the formation of two bulk phases separated by a surface tension controlled interface. Instead, the composition varies continuously from the center to the boundary of the arena, and the radial density profiles exhibit a self-similar form across different system sizes (see Supplementary Material).

Although circular arenas provide an efficient geometry for establishing radial demixing, they are not well suited for the continuous collection of the separated particle populations. A practical self-sorting device, therefore, requires a channel geometry that preserves the underlying demixing mechanism while directing the two particle fractions toward distinct outlets.
\begin{figure*}[!t]
\centering
\includegraphics[width=\textwidth]{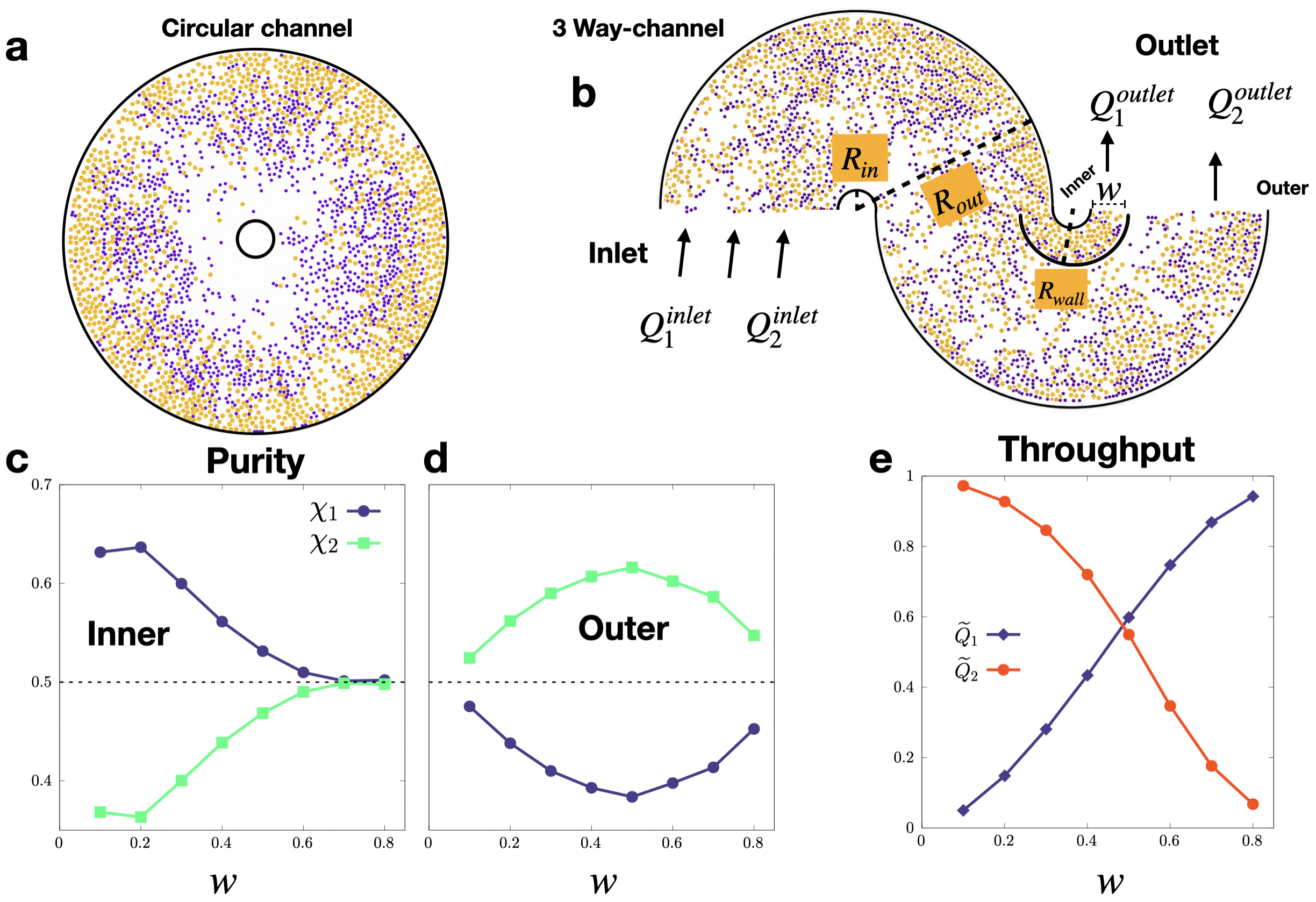}
\caption{From circular channels to continuous particle collection.
\textbf{a.} Circular annular channel used to generate radial demixing between fast and slow particles.
\textbf{b.} Three-way sorting geometry obtained by combining two shifted semicircular channels and introducing a separating wall that directs the demixed populations towards distinct inner and outer outlets. The wall position is parametrized by $w=(R_{\rm wall}-R_{\rm in})/(R_{\rm out}-R_{\rm in})$.
\textbf{c.} Steady-state particle fractions $\chi^{\rm in}_{1}$ and $\chi^{\rm in}_{2}$ at the inner outlet as a function of $w$. The purity of fast particles is maximal around $w\simeq0.2$.
\textbf{d.} Corresponding particle fractions $\chi^{\rm out}_{1}$ and $\chi^{\rm out}_{2}$ at the outer outlet, with maximal purity of slow particles around $w\simeq0.5$.
\textbf{e.} Normalized throughput of fast particles recovered at the inner outlet, $\tilde{Q}_{1}$, and slow particles recovered at the outer outlet, $\tilde{Q}_{2}$, as a function of $w$. Increasing $w$ enhances the recovery of fast particles while reducing that of slow particles, revealing the trade-off between outlet purity and particle recovery.}
\label{Fig4}
\end{figure*}

To continuously collect the sorted particles, the demixed populations must ultimately be directed towards separate outlets. This requirement introduces an important geometrical constraint: curvature plays a key role in promoting demixing in colloidal flocks of Quincke rollers~\cite{PhysRevLett.131.178304}. To determine how the sorting mechanism depends on confinement geometry, we therefore designed an annular channel by introducing an impenetrable circular region of radius $R_{\mathrm{in}}$ at the centre of the original circular arena of radius $R_{\mathrm{out}}$. Varying the ratio $R_{\mathrm{in}}/R_{\mathrm{out}}$ allows us to systematically tune the relative curvature of the inner and outer boundaries. Experiments and simulations were performed for a 50:50 binary mixture with $a_2/a_1=0.75$, $v_2/v_1=0.8$, and packing fraction $\phi=0.2$.

Both experiments and simulations reveal a pronounced dependence of the steady-state organization on this geometrical ratio. At small $R_{\mathrm{in}}/R_{\mathrm{out}}$, the two species remain radially demixed, whereas increasing $R_{\mathrm{in}}/R_{\mathrm{out}}$ progressively suppresses demixing and eventually produces a mixed steady state (Fig.~\ref{Fig3}a). We quantify this transition using the time-averaged steady-state demixing parameter, $\langle\gamma\rangle_t$, shown in Fig.~\ref{Fig3}b. The demixing parameter decreases monotonically with increasing $R_{\mathrm{in}}/R_{\mathrm{out}}$, confirming that radial segregation becomes progressively weaker as the relative curvature of the two boundaries decreases. 

Therefore, demixing is not determined solely by the curvature of the outer boundary, although finite-size effects remain present. Instead, the relative geometry of the two confining boundaries, captured by $R_{\mathrm{in}}/R_{\mathrm{out}}$, provides the dominant control parameter for maintaining radial segregation in an annular channel. Efficient sorting, thus, requires sufficiently small $R_{\mathrm{in}}/R_{\mathrm{out}}$, establishing a direct geometrical design principle for translating curvature-induced demixing into a continuous particle-sorting device.

\begin{figure*}[!t]
\centering
\includegraphics[width=\textwidth]{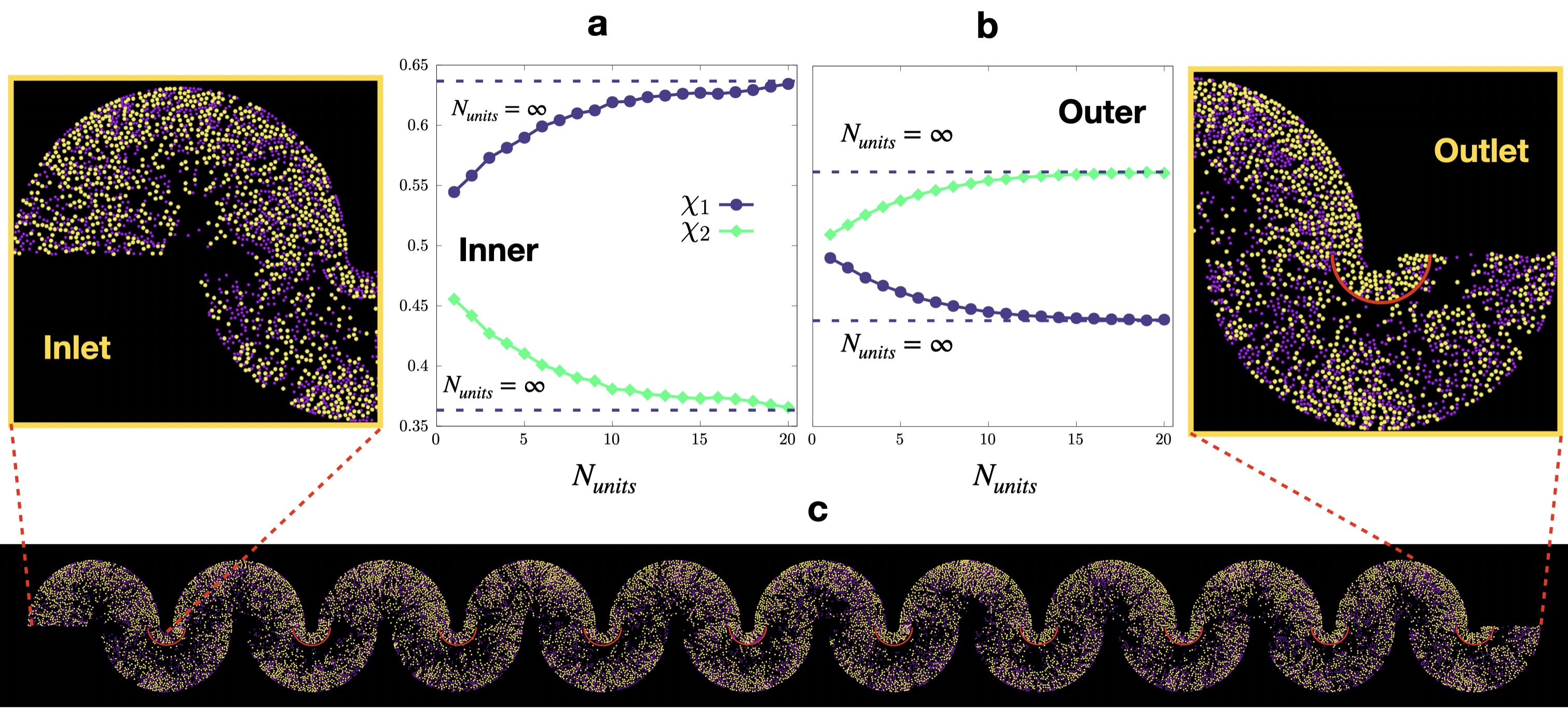}
\caption{From periodic sorting to a finite-size device. \textbf{a.} Steady-state particle fractions measured at the inner and outer outlets as a function of the number of three-way sorting units, ($N_{\rm units}$). Dashed lines indicate the corresponding values obtained in the infinite-channel limit. The inner-outlet composition converges after approximately ($N_{\rm units}\simeq20$), whereas the outer outlet reaches its asymptotic value after about ($N_{\rm units}\simeq12$). \textbf{b.} Schematic of the finite sorting device composed of successive three-way channel units, showing the progressive segregation of the two particle species along the channel. \textbf{c.} Reinjection boundary condition used to mimic continuous operation of a finite device. Particles leaving through either outlet are reinjected at a random position across the inlet while retaining their velocity. Starting from a mixed population at the inlet, repeated passage through the sorting units produces a pronounced downstream demixing, with faster particles preferentially localized near the inner wall and directed towards the corresponding outlet.
}
\label{Fig5}
\end{figure*}

We next use numerical simulations to address a second design challenge: connecting a single inlet to two separate outlets while preserving the demixing generated by the curved channel. Starting from the circular geometry of Fig.~\ref{Fig4}a, we cut the channel into two semicircular sections, translate them relative to one another, and reconnect them as illustrated in Fig.~\ref{Fig4}b. Because the direction of radial demixing reverses between the two semicircular sections, particles separated in the first half would otherwise be driven towards one another and remix in the second. We therefore introduce a separating wall at the junction between the two sections, which preserves the spatial separation established upstream and directs the two populations towards distinct outlets. This construction results in the three-way geometry shown in Fig.~\ref{Fig4}b.

The performance of this geometry depends on the position of the separating wall. We parameterize its radial location by
\begin{equation}
w=\frac{R_{\rm wall}-R_{\rm in}}{R_{\rm out}-R_{\rm in}},
\end{equation}
where $R_{\rm wall}$ is the radial distance of the wall from the center of the corresponding semicircular section. Guided by the results of the previous section, we choose a compact and strongly curved channel, for which demixing is both rapid and pronounced. We therefore set $R_{\rm out}=50$ and $R_{\rm in}/R_{\rm out}=0.1$, while keeping all remaining simulation parameters unchanged from the previous section.

To first identify strategies for optimal demixing, we use simulations to consider the idealized limit of an infinite sequence of sorting units. To realize this limit, we impose periodic boundary conditions that reconnect the two outlets to the inlet. Particles leaving either outlet are therefore reinjected at the inlet, allowing the system to reach a statistically stationary state while maintaining a constant particle population. Within this periodic geometry, we systematically vary $w$ and quantify the resulting sorting performance from the particle composition at the two outlets. Figure~\ref{Fig4}c reports the corresponding steady-state fractions $\chi^{\rm in}_{1}$ and $\chi^{\rm in}_{2}$ at the inner outlet, whereas Fig.~\ref{Fig4}d shows $\chi^{\rm out}_{1}$ and $\chi^{\rm out}_{2}$ at the outer outlet, with
$\chi^{\rm in}_{1}=1-\chi^{\rm in}_{2}$ and
$\chi^{\rm out}_{1}=1-\chi^{\rm out}_{2}$.

The outlet composition depends strongly on the wall position. For collection at the inner outlet, the purity varies non-monotonically with $w$, reaching an optimum value around $w\simeq0.2$, where $\chi_1$ is maximal and $\chi_2$ is minimal (Fig.~\ref{Fig4}c). Beyond this point, the purity decreases monotonically with increasing $w$, and the collected population becomes progressively more mixed. A similar non-monotonic dependence is observed for collection at the outer outlet, although the optimal purity is shifted to $w\simeq0.5$ (Fig.~\ref{Fig4}d). For larger $w$, the purity again decreases monotonically, approaching a mixed state.
\begin{figure*}[!t]
\centering
\includegraphics[width=\textwidth]{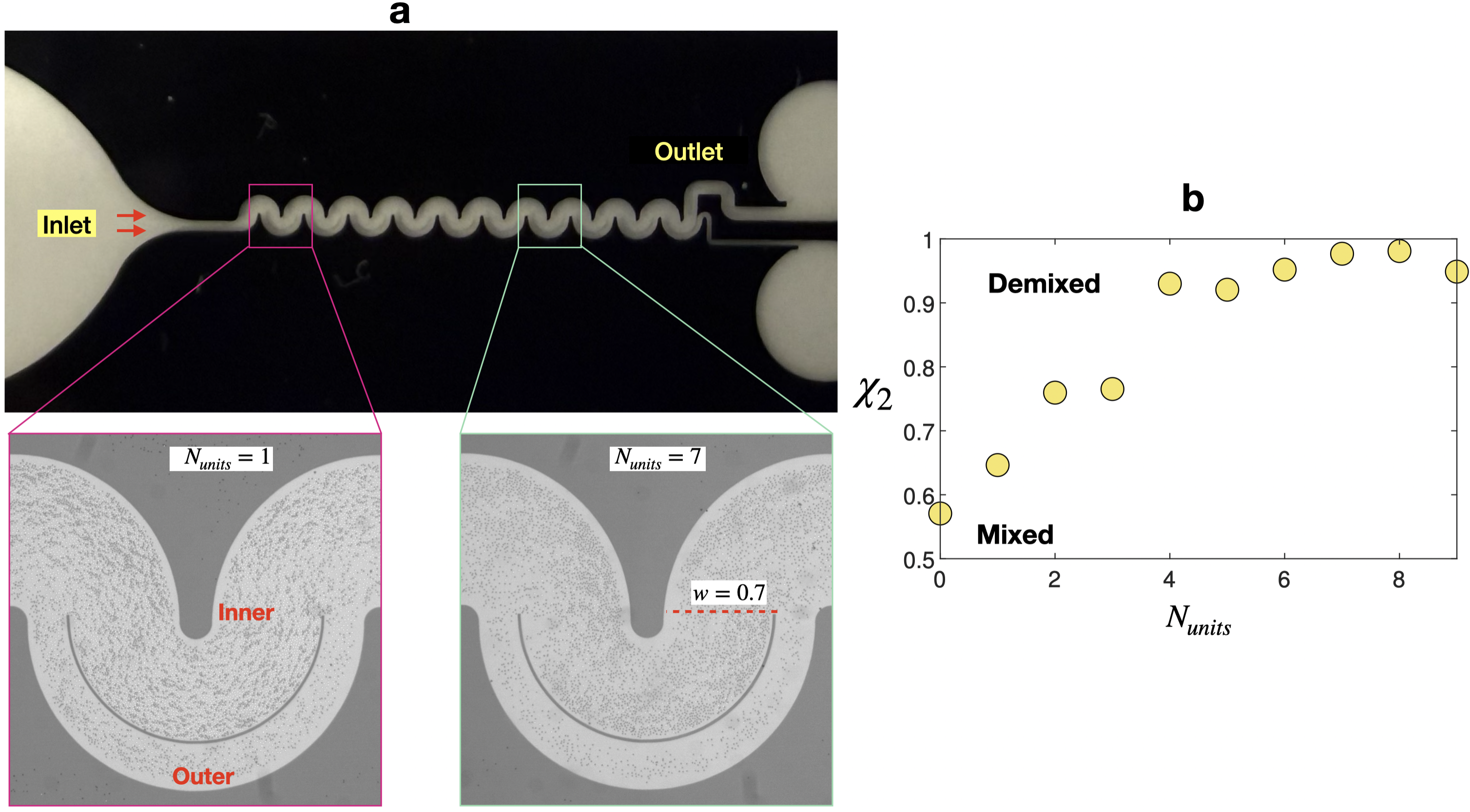}
\caption{\textbf{Demixing in snake-like channels.}
\textbf{a.} Top: Geometry of the self-sorting device ($w=0.7$). Insets: Experimental snapshots of the operating device. 
\textbf{b.} Density fraction of the slow, $7\,\rm\mu m$-diameter PS particles, in the outer region as a function of the length of the channel expressed in number of elementary units.
Demixing reaches high purity in about 10 units.
}
\label{Fig6}
\end{figure*}
Purity alone does not fully characterize the performance of the sorter, because a highly pure outlet may recover only a small fraction of the target species. We therefore also quantify particle recovery through the normalized throughput
\begin{equation}
    \tilde{Q}_{k}
    =
    \frac{Q^{\rm outlet}_{k}}{Q^{\rm inlet}_{k}},
\end{equation}
where $Q^{\rm inlet}_{k}$ is the incoming flux of species $k$ and $Q^{\rm outlet}_{k}$ is the flux of the same species recovered at its designated outlet. Fast particles (species 1) are collected at the inner outlet, whereas slow particles (species 2) are collected at the outer outlet. Accordingly, $\tilde{Q}_{1}$ and $\tilde{Q}_{2}$ represent the fractions of incoming fast and slow particles successfully recovered at their respective target outlets.

Figure~\ref{Fig4}e shows that the two throughputs vary oppositely with the wall position. The recovery of fast particles at the inner outlet, $\tilde{Q}_{1}$, increases monotonically with $w$, reaching its largest value when the separating wall approaches the outer boundary. Conversely, the recovery of slow particles at the outer outlet, $\tilde{Q}_{2}$, decreases monotonically with $w$ and is therefore largest when the separating wall lies close to the inner boundary. The two curves intersect around $w\simeq0.5$, where the two species are recovered with comparable efficiency.

Comparison with the corresponding purity curves reveals a trade-off between purity and recovery: wall positions that yield a highly enriched outlet do not necessarily collect the largest fraction of the target species. The optimal wall position must therefore balance these two complementary measures, with purity quantifying the composition of the collected fraction and $\tilde{Q}_{k}$ quantifying the fraction of the available target population that is recovered.

So far, we have characterized the sorting performance in terms of steady-state purity and throughput using periodic boundary conditions. Although this approach provides access to the asymptotic behavior of an effectively infinite sequence of sorting units, it does not directly represent the operation of a finite device. In practice, the available space is limited, and particles traverse only a finite number of three-way channels before being collected at the outlets. This raises an important practical question: what is the minimum number of sorting units required for the outlet composition to approach that obtained in the infinite-channel limit?

To address this question, we introduce a modified reinjection rule at the virtual boundary connecting the outlets to the inlet. When a particle crosses an outlet, it is reinjected at a random position along the inlet cross-section while retaining its velocity (Fig.~\ref{Fig5}c). This procedure mimics the continuous supply of particles in an experiment while eliminating the spatial correlations generated by directly reconnecting each outlet to the inlet. The resulting steady state therefore allows us to monitor how an initially mixed population progressively demixes as it passes through successive three-way sorting units.

The simulations show that particles enter the device in a mixed state and progressively segregate as they propagate downstream. After traversing several sorting units, a pronounced radial organization develops, with the faster particles preferentially localized near the inner wall and consequently directed towards the corresponding outlet (Fig.~\ref{Fig5}c). We quantify this convergence by measuring the particle fractions $\chi_1$ and $\chi_2$ at the inner and outer outlets as a function of the number of sorting units, $N_{\rm units}$. For the inner outlet, the steady-state fractions progressively approach their infinite-channel values and are essentially converged by $N_{\rm units}\simeq20$ (Fig.~\ref{Fig5}a). This demonstrates that the demixing obtained in the infinite-channel limit can be reproduced with a finite, experimentally realizable device.

For the outer outlet, convergence is reached more rapidly, at approximately $N_{\rm units}\simeq12$. The corresponding compositions nevertheless remain closer to the mixed-state value than those at the inner outlet. This asymmetry reflects our choice of wall position, which was optimized to maximize the purity of the faster particles collected at the inner outlet rather than that of the complementary fraction. These results provide a direct design criterion for a finite sorting device: a few tens of three-way units are sufficient to reproduce the asymptotic sorting performance while retaining a compact, experimentally accessible geometry.
\section{Experimental self-sorting device}
\label{sec:experiment_device}
From this numerical exploration and insights gained, we now have all the information to realize an experimental self-sorting device. We develop a snake-like channel composed of successive curved sorting units, as shown in Fig.~\ref{Fig6}a. The geometry is chosen to preserve the curvature-induced demixing mechanism while remaining compatible with the finite area available on the microfluidic chip. Guided by the simulations, we set $R_{\rm in}/R_{\rm out}=0.1$, $R_{\rm out}=1~\mathrm{mm}$, $N_{\rm units}=10$, and place the
position of the separating wall at $w=0.7$. These parameters provide a compact realization of the sorting geometry while allowing the particle composition to evolve progressively as the suspension travels downstream.

The experiments are performed using a binary mixture of fluorescent polystyrene particles with diameters $d_{\rm black}=7~\mu\mathrm{m}$ and $d_{\rm white}=10~\mu\mathrm{m}$, prepared in approximately equal proportions. A continuous suspension of the two species is injected through the inlet using a low-pressure syringe pump, providing a sustained flux of initially mixed particles through the device. The particles then traverse the sequence of curved units before reaching the outlet region. Further details of the experimental preparation and driving conditions are provided in the Supplementary Material.

As the particles propagates along the channel, the initially mixed population progressively develops the radial organization identified in the simulations. After only a few sorting units, the two species become increasingly segregated, and the separating wall converts this spatial organization into selective particle collection. For the chosen geometry, the smaller $7~\mu\mathrm{m}$ and slower particles are preferentially directed toward the outer region of the channel and subsequently collected at the corresponding outlet.

We quantify this progressive sorting by measuring the fraction $\chi_2$ of $7~\mu\mathrm{m}$ smaller (slower) particles in the outer region as a function of the distance traveled through the device, expressed in the number of sorting units. As shown in Fig.~\ref{Fig6}b, $\chi_2$ increases systematically downstream and approaches unity after approximately seven units. Thus, repeated passage through the curved geometry progressively amplifies the compositional imbalance, transforming an approximately equimolar inlet suspension into a highly enriched particle stream at the outlet. This experimental realization demonstrates that curvature-induced collective demixing can be translated from circular confinement into a finite continuous-flow device, enabling robust and autonomous sorting of active particles.

\section*{Conclusions}
We have demonstrated that collective demixing in binary mixture of Quincke rollers can be harnessed as a physical mechanism for continuous sorting. Combining experiments with particle-based simulations, we showed that differences in speed are sufficient to generate robust radial segregation under curved confinement. The faster species preferentially occupies the outer region of the flock, and reversing the relative propulsion speeds reverses the spatial organization, identifying motility difference as the primary control parameter for the demixing.

We further established confinement geometry as a key ingredient for controlling the efficiency and timescale of segregation. Smaller circular confinements accelerate demixing without substantially compromising its steady-state strength, while increasing $R_{\rm in}/R_{\rm out}$ in annular channels progressively suppresses segregation. Guided by these observations, we designed via numerical simulation a three-way channel in which a separating wall converts radial organization into distinct particle streams. The wall position controls the balance between outlet purity and particle recovery, providing a simple geometrical parameter for tuning sorting performance.

Finally, from the numerical exploration and insights gained, we  created an experimental self-sorting device. We develop a snake-like channel composed of successive curved sorting units and demonstrated experimentally that an initially mixed population of rollers progressively segregates as it propagates through successive curved units, reaching a highly enriched state downstream. These results establish a direct connection between demixing in active systems and functional particle transport, showing that spontaneous self-organization can be engineered into a practical continuous-flow sorting mechanism.

More broadly, this work opens new possibilities for designing technologies that exploit collective segregation, sort, and transport microscopic components for specific tasks. Future devices could harness dynamical properties such as motility and collective interactions together with curved confinement geometry to control particle organization and transport. Such strategies may provide new routes toward active microfluidic platforms for materials processing, particle selection, and biomedical applications, including the controlled separation of motile cells or microorganisms.

\begin{acknowledgments}
We thank Samadarshi Maity for help with the experiments. We thank Joshua H. K. Saldi for insightful discussions.
 
\end{acknowledgments}
\bibliographystyle{apsrev4-2}
\bibliography{references}
\end{document}